\documentclass[aps,prl,showpacs,preprint]{revtex4}
\usepackage{graphicx}
\usepackage{epsfig}

\begin{document}


\input{epsf.tex}
\epsfclipon
\newcommand{\beq}{\begin{equation}}
\newcommand{\eeq}{\end{equation}}

\title{Time-Dependent Trapping of Solitons in Bose-Einstein Condensates}

\author{H.E. Nistazakis$^1$, D.J. Frantzeskakis$^1$, N. Brouzakis$^1$, F.K. Diakonos$^1$,
P. Schmelcher$^2$ and J. Schmiedmayer$^3$} 

\address{$^1$  Department of Physics, University of Athens, GR-15771 Athens, Greece\\
$^2$  Theoretische Chemie, Physikalisch-Chemisches Institut, Im Neuenheimer Feld 229,
69120 Heidelberg, Germany \\
$^3$ Physikalisches Institut, Philosophenweg 12, 69120 Heidelberg, Germany}

\begin{abstract}
We study the influence of a time-dependent potential on the
motion of solitons in a quasi one-dimensional Bose-Einstein condensate
by solving the corresponding Gross-Pitaevskii equation.
For a suitable choice of the external potentials as
well as the initial soliton characteristics time-dependent trapping of the
soliton in a prescribed subarea of the condensate can be achieved.
Adiabatic perturbation theory is shown to work remarkably well for
large switching on times of the trapping potential and allows to perform
a detailed study of the degree of trapping in the complete phase space of the
soliton center. A remarkable spiral pattern of the degree of trapping as a function
of the soliton characteristics is observed and explained. 
\end{abstract}


\maketitle


\section{Introduction}

Since the experimental discovery of Bose-Einstein Condensation (BEC) in
dilute alkali metal vapours in 1995 \cite{Kett97} this field has experienced
an enormous increase of interest (for recent reviews see refs.\cite{RMP1,
RMP2}). Indeed, the perspectives are very promising and range from coherent
matter wave optics such as atom lasers, interferometry or atom lithography to
precision measurements and quantum information processing. External static
electric and magnetic or electromagnetic fields are of equal importance to
the production as well as manipulation of the condensed phase. Using
different traps it is nowadays possible to produce in particular low i.e.
(quasi-) one- and two-dimensional condensates \cite{Ketterle96}.

One of the most promising ideas for the manipulation of coherent
matter waves is the so-called atom chip which consists of a network of charge
and current carrying elements, such as wires, on, for example, a semiconductor surface
\cite{Muel} (for a comprehensive review see R. Folman et al. \cite{Folman}).
The electric and magnetic fields generated by the corresponding
charge distributions and currents can then, together with homogeneous
external fields, be employed to guide and control the coherent atomic ensemble which
is moving a few micron above the surface. In particular the different field
configurations offer the possibility to create microtraps, waveguides
and other atom-optical devices on a single atomic chip. Very recently the
first BEC on such an atom chip has been prepared \cite{Haen,Ott,Schneider,Leanhardt} using
microscopic traps. Subsequently the condensate has been coupled to an
atomic conveyer belt \cite{Haen2} which is used to transport the condensed
cloud nondestructively over a macroscopic distance parallel to the chip
surface \cite{Haen}. 

According to the above optics with atoms being either in the thermal or
condensate phase is now well-established. An important question posed 
here is whether one can do optics with excitations of the condensate.
Specifically we will study the possibility to control
the motion of nonlinear excitations of the condensate,
in general, and in particular of solitons.
Experimentally there exist several quantum-phase-engineering
techniques to generate dark solitons in a Bose-Einstein condensate
(see, e.g., ref. \cite{Denschlag00}) which, according to theoretical
predictions \cite{Rein}, oscillates in the BEC. 
The question then arises how one could influence or even guide their motion, e.g. by modifying 
the above mentioned domains, so as to transfer ``information'' via solitons
in a controllable manner. Here we provide a step in this direction by investigating the possibility
of trapping the motion of a soliton in a prescribed subarea of the condensate via a time-dependent potential.
The time-dependent change of the potential can, for example, be implemented on atom 
chips by switching currents flowing in the corresponding wires.
As we shall see the degree of trapping $m_t$ in the subarea, that is a measure for the strength
of the confinement of the motion of the soliton, shows a twofold dependence.
First it depends on parameters such as the switching time of the potential.
Second it turns out that, for fixed parameters, the degree of trapping $m_t$ is a function
of the phase space of the soliton center showing a beautiful spiral pattern. 

\section{Solitons and the Gross-Pitaevskii Equation}

The above condensates consist of dilute ensembles of weakly
interacting alkali atoms. In spite of its weakness the interaction is responsible for many of the
properties of the condensate. In particular the diluteness at the nano-Kelvin scale
implies that the dominant s-wave scattering of binary collisions provides a
scattering length $a$ which is much smaller than the mean interatomic
spacing. As a consequence it is for many purposes a good approximation
to introduce a contact potential as an effective interaction. In a mean-field
description this yields the Gross-Pitaevskii equation (GPE)
(see ref.\cite{RMP1} and references therein) 
\begin{equation}
i \hbar  \frac{\partial}{\partial t} \Psi(\vec{r},t)=
\left( - \frac{\hbar^2}{2m} \nabla^2 + V_{ext} + g \vert
\Psi(\vec{r},t)\vert^2\right) \Psi(\vec{r},t)
\label{eq:eq0}
\end{equation}
where $m$ is the mass of a single atom, $g=\frac{4\pi\hbar^2 a}{m}$ is the coupling constant
and $V_{ext}$ represents an external potential. In our case $V_{ext}$ will contain both the external time-independent
potential which provides the overall confinement of the BEC as well as a time-dependent part to be specified below.

Due to the nonlinearity occuring in eq.(\ref{eq:eq0}) the GPE is capable of
describing energetically low-lying collective excitations of the
condensate such as solitons. For repulsive atom-atom interaction ($g > 0$),
which is the situation investigated here, only dark solitons occur.
Solitons are well-known to play an important role in traditional nonlinear
(fibre) optics. In the case of coherent matter waves the existence of solitons has been
demonstrated recently followed by a number of investigations on their
properties \cite{Rein,Feder00,Brand}.

Eq.(\ref{eq:eq0}) provides a mean-field description of the condensate and its
excitations in three space dimensions. However, present day experiments allow to
prepare in particular low-dimensional condensates of (quasi-)two or one dimensional
\cite{Ketterle96} character showing very interesting thermodynamical as well as microscopic properties.
In the case when the confinement for two of the three spatial dimensions is much stronger than in the third
dimension,  the GP equation can be reduced to an effective quasi-$(1+1)$-dimensional GP equation \cite{GP1dper}.
For repulsive inter-atomic interactions, the latter equation can be
expressed in the following dimensionless form,
\begin{equation}
iu_{t}+\frac{1}{2}u_{xx}-|u|^{2}u=V_{ext}(x)u,
\label{gp}
\end{equation}
where the spatial coordinate $x$ and time $t$ are normalized to
the harmonic oscillator length  $\alpha_{\perp }=\sqrt{\hbar
/m\omega _{\perp } }$ and oscillation period,  $1/\omega _{\perp
}$ respectively. The frequency $\omega_{\perp}$ belongs to the two
dimensions with strong confinement. The field $u$
describes the macroscopic wave function $\Psi$ of the condensate,
according to the following scaling relation
\begin{equation}
\Psi (x,t) = \left( \frac{m \omega_{\perp}}{4 \pi a \hbar}\right)^{1/2} u(x,t)
\end{equation}

The external potential $V_{ext}$ is decomposed into a
time-independent part $V_1$ which ensures the overall confinement for all times
and an explicitly time-dependent part according to the following appearance

\begin{equation}
V_{ext}(x,t)=V_1(x)+ f(t-t_o) V_2(x)
\label{eq:eq2}
\end{equation}

The additional potential $V_2$ is turned on adiabatically. The major change of the potential occurs within
the time interval $[-t_c/2+t_o, t_c/2+t_o]$ where $t_o$ is the characteristic switching on time and $t_c$ is the
duration of the switching on process. Asymptotically ($t \to \infty$) the full potential $V_1 + V_2$ has been
turned on. In the case under consideration the function $f(t)$, which plays the role of a switch, is chosen to be
\begin{equation}
f(t)=\frac{1}{2}\left(1 + \tanh(\frac{t-t_0}{t_c})\right)
\label{eq:eq3}
\end{equation}
In the limit $t_c \to 0$ $f(t)$ leads to the step function $\theta(t)$ i.e. we encounter an instantaneous
switching on process 
\begin{equation}
V_{ext}^{(0)}(x,t) = \left\{\begin{array}{c} V_1(x), \hspace*{1.4cm} t<t_o \\
                                           V_1(x)+V_2(x),~~~t \geq t_o
\end{array} \right.
\end{equation}

It is important to note that the role of the additional potential is, as we
will see, to confine the motion of the solitons within a certain spatial region, which can be varied almost
arbitrarily by the particular choice of $V_2(x)$. There are two possibilities concerning the effect of $V_2(x)$
on the shape of the initial trapping potential $V_1(x)$, namely to add either a barrier or a well to the
confining potential. In the former case the spatial extension of the ground state wave function is
not affected significantly; however, in the latter case, the spatial extension of the new ground state
is clearly reduced leading, in the extreme cases of very deep and narrow wells, to the collapse of the condensate. 
We will therefore consider only the case where $V_2(x)$ adds a barrier to the confining potential $V_1(x)$.

\section{Computational Approach}

The GPE in eq.(\ref{gp}) is solved numerically using the
Split-Step Fourier method \cite{SSF}. In the numerical
simulations we set up a solitary excitation for the initial time $t=0$
using the following Ansatz for the wave function of the soliton \cite{Frantzeskakis}
\begin{eqnarray}
u(x,0)&=&\left[\sqrt{\mu} - \frac{1}{2 \sqrt{\mu}} U(x,0)\right]
\left( \cos \phi(0) \tanh \xi + i \sin \phi(0) \right) 
\nonumber \\
\xi&=& \left(x - x_o (0) \right) \cos \phi(0)
\label{eq:eq4}
\end{eqnarray}
where $\mu$ is the chemical potential of the BEC,  
$x_o(0)$ is the position of the soliton center
at $t=0$ and $\phi(0)$ the corresponding soliton phase angle
($\vert \phi(0) \vert < \frac{\pi}{2}$). Assuming that the soliton trapping
potential $V_2$ is switched on much later than $t=0$, i.e. $t_o \gg 0$,
we can take the function $U(x,0)$ in eq.(\ref{eq:eq4}) equal to the BEC
trapping potential $V_1(x)$. The expression (\ref{eq:eq4}) is based on the results of
adiabatic perturbation theory for dark solitons \cite{Adper}, which can be applied in the
case where $V_1(x)$ varies slowly on the soliton scale, and turns out to be a very good
estimate for the initial soliton wave function \cite{Frantzeskakis}.
Solving eq.(\ref{gp}) numerically, we then obtain the evolution of the soliton on top of the
condensate. The fact that the initial form of the soliton (\ref{eq:eq4}) is an approximation to the
exact soliton wave function introduces, besides the standard numerical propagation error, 
an additional source of inaccuracy
with respect to the soliton dynamics. However, as can be seen in the following, 
the time evolution of the Ansatz
(\ref{eq:eq4}) is very close to the expected solitary wave propagation.
If the switching on time interval $t_c$ is large then the potential
varies slowly with time and the adiabatic perturbation theory can be used
to estimate, at least approximately, the soliton + BEC dynamics. In fact we will see later
that the critical value for $t_c$ below which the 
adiabatic perturbation theory breaks down is $t_c \approx 10$.
Particularly, the equations describing, within adiabatic perturbation theory (for $t_c > 10$), 
the motion of the soliton center \cite{Rein,Frantzeskakis} are given by:
\begin{equation}
m_{eff} \frac{d^2 x_o}{d t^2} = - \frac{\partial U}{\partial x_o}~~~;~~~
m_{eff}=\frac{2 \sqrt{\mu}}{\cos^2 \phi(0)}
\label{eq:eq5}
\end{equation}
and are asymptotically valid, i.e. for $t \to -\infty$ or $t \to \infty$ where
the time dependence of the external potential can be ignored. For slowly varying 
switching on function $f(t)$ ($t_c > 10$) one can
take for the effective potential $U = V_{ext}$ where $V_{ext}$ is the total
external potential. 
Equations (\ref{eq:eq5}) can, in general, be solved
numerically to obtain the trajectory of the
center of the soliton. The results obtained by the direct numerical integration of the GPE are in a fairly good agreement with the ones obtained by the numerical integration of eqs.(\ref{eq:eq5}). Therefore, the latter can safely serve as a guide for the choice of the appropriate 
initial conditions for potential trapping used in the much more expensive numerical propagation of the GPE.

Having fixed the form of the external potential the only free
parameters entering in the solution of eq.(\ref{gp}) are the initial
values $x_o(t=0),\phi(t=0)$ of the position of the center and the
phase of the soliton, respectively. After switching on the potential $V_2(x)$,
the possible trapping of the dark soliton in a prescribed region of the condensate depends
strongly on $x_o(0),\phi(0)$. The amplitude, and therefore also the
velocity, of the soliton are determined through $\sin\phi(0)$. Large values for
the amplitude lead to a slowly moving soliton. If the soliton is moving very slowly
the time needed to perform a full oscillation within the condensate is
comparable with the lifetime of the condensate itself: In this case trapping
is not an issue. If the amplitude of the soliton is small, the velocity is large and
the trapping of the soliton becomes impossible. 
One could think of avoiding the loss of trapping by using a more pronounced
external potential. However this leads in general to a highly unstable condensate 
wave function.
As a consequence there are several competing factors which determine
the evolution of the composite system (soliton + condensate) and its
properties: (a) the initial conditions of the solitonic excitation ($x_o(0),
\phi(0)$) controlling the evolution of the soliton, (b) the form of the
potential determining the shape of the condensate and (c) the
duration $t_c$ of the switching on process of the soliton trapping potential
$V_2$. If the potential is turned on suddenly the BEC becomes
very unstable and there is no initial configuration for the soliton leading
to trapping before the BEC is destroyed. Therefore we expect to have trapping of the soliton,
which can be clearly observed within the BEC lifetime, only for restricted
values of $x_o,\phi(0)$ and $t_c$ as well as special forms of the soliton trapping
potential $V_2$.

\section{General Aspects of Trapping}

Before proceeding with the discussion of our numerical results let us define
the measure $m_t$ for the degree of trapping for a given trajectory of the center of a soliton 
in a subarea of the BEC
\begin{equation}
m_t=\frac{A^{(-)}-A^{(+)}}{A^{(-)}} 
\end{equation}
where $A^{(-)}, A^{(+)}$ are the amplitudes of the oscillation of the center of the soliton for
times $t\ll t_o$ and $t\gg t_o$, respectively. We will investigate
soliton trapping for two different cases of the external potential $V_{ext}$.
The condensate trapping potentials $V_1(x)$ possesses in both cases a single minimum.
After turning on the soliton trapping potential the total potential
$V_{ext}$ will exhibit two minima for the external potential $V^{(2)}_{ext}$ and 
correspondingly three minima for $V^{(3)}_{ext}$ (see figs. 3 and 5, respectively).
In order to reduce the number of free parameters in our study we fix the values for
the parameters of the external potential such that the BEC ground state
matches well the dimensions of the experimentally prepared condensates. 

Let us first consider the external potential 
\begin{equation}
V^{(2)}_{ext}(x,t)=a x^4 + f(t-t_o)(\beta x^2+ \gamma)
\label{eq:eq6}
\end{equation}
where $\gamma=0.2$, $\beta =-0.001$, $\alpha =1.25 \cdot 10^{-6}$, $t_c=40$
and $t_o=600$.  Having fixed the external potential we remain with the free parameters $x_o(0)$ and
$\phi(0)$ which determine the solitonic excitation. First we
calculate the degree of trapping $m_t$ for a $100 \times 100$ grid in the
($x_o(0),\phi(0)$) space. The results of these calculations are shown in the
contour plot of Fig.~1. They are obtained employing adiabatic perturbation theory.
For a subset of the gridpoints we have also calculated, as a test,
$m_t$ using the full GPE propagation. The results turn out to differ at most by $20\%$ for this value of
$t_c$ (The integration of the GPE is computationally too expensive in order to obtain a 
well-resolved $sin(\phi_o)-x_o$ plot in Fig.1).
We observe a spiral pattern which determines the region with significant trapping 
($m_t \geq 40\%$) of the soliton in the condensate. There are regions with
trapping of the order of $100\%$ depicted by the white areas in the arms of the spiral pattern.
We have repeated our calculations for larger as well as for smaller $t_c$. For increasing
$t_c$ (we have chosen $t_c=70$) the arms of the spiral become thicker
and finally they meet each other forming a connected almost rectangular region
where trapping occurs. Decreasing $t_c$ ($t_c=15$) we observe that the arms
of the spiral shrink more and more and the trapping region tends to disappear.

The origin of this spiral pattern of the trapping measure $m_t$ can be understood 
qualitatively by recalling that the switching on process is accompanied by
an energy transfer to the soliton. The soliton center trajectory moves for $t \ll t_o$,
i.e. before the trapping potential is turned on ($V=V_1$), on an closed phase curve 
in the phase plane $(x_o, sin^2 \phi)$. During the switching on process this phase curve is
deformed. In case of strong trapping for $t \gg t_o$ ($V=V_1+V_2$) the resulting phase
curve is a closed phase curve again but with a significantly smaller size. In other words the available
phase space volume for the motion of the soliton center has shrinken significantly due to
the trapping. The transition between these two phase curves occurs via an inward spiral which
represents the natural transient between the two closed curves of different size.
Quantitatively the spiral pattern can be understood by calculating the energy transfer
in e.g. the limit of a sudden switching on and imposing the final condition of a strong
confinement i.e. trapping.

\section{Reflection and Trapping of Solitons}

The basic mechanism leading to the soliton trapping is the reflection of the soliton
at the walls of the external potential. At the reflection point of the soliton center
its kinetic energy vanishes and the soliton center therefore possesses only potential
energy. During this reflection process the phase of the
soliton is changed dramatically. Such a reflection process is monitored in Fig.~2a where three different
time instants (before, at and after the reflection on the central hump of the external potential)
are depicted in the trajectory
of the soliton center obtained by numerical integration of the GPE. 
In Fig.~2(b,c,d) we show the phase function of the soliton
for these three time instants, as well as the corresponding probability density 
$\vert u \vert^2$. In particular, for $t=982$ (before the reflection), Fig.~2b shows 
the soliton phase as a function of $x$, which is characterized by a smooth jump from a region of low phases
(left of the soliton center) to a region of high phases (right of the soliton center).
As the soliton approaches the barrier its depth increases and the phase function 
becomes steeper. As a result, at the reflection point shown in Fig.~2c ($t=1041$),
the phase jump becomes step-like, i.e. discontinuous and after the reflection
(see Fig.~2d for $t=1104$) the regions of low phases and high phases are interchanged while the change
of the phase across the soliton center becomes smooth again. These three snapshots
are calculated using the same parameters for the external potential as in Fig.~1.

In Fig.~3a we show snapshots of the soliton+BEC dynamics for twelve subsequent time instants
using the initial conditions $(x_o(0),\sin\phi(0))=(-5,\sqrt{0.15})$ and $t_c=40, t_o=600$.
The soliton trapping is clearly visible for times $t > 600$ for which the motion exclusively takes place in the right well
of the potential. The small fluctuations due to radiation of 
the BEC can be well distinguished from the solitonic excitation up to times 
$t \approx 1200$. For significantly larger times the size of the fluctuations becomes
of the order of the soliton amplitude finally leading to the decay of the condensate. 
A rough estimation of the trapping degree in this
case gives $m_t \approx 0.6$. This can be also seen in Fig.~3b where the 
corresponding trajectory of the soliton center (solid line) is shown. In the same plot 
we present also the results of adiabatic perturbation theory for a comparison (dashed line). 
It is readily seen that the results of perturbation theory fit very well the
numerical calculations. The relative error in the present case in the amplitude of the soliton oscillations is less than $30\%$. We have repeated our calculations for two other
choices of the switching on parameter $t_c$. The results of the calculations are given
in Fig.~4(a,b). In Fig.~4a we show the trajectory of the soliton center for the potential
(\ref{eq:eq6}) using $t_c=15$. Presented are both the GPE integration results (solid line)
as well as the adiabatic perturbation theory results (dashed line). It is clearly seen that
in this case (small $t_c$) the discrepancy between perturbation theory and exact numerical
integration is much larger than for $t_c=40$ in figure 3.
The opposite behaviour is observed in Fig.~4b where we present analogous results
for the case $t_c=70$.   

The above shape of the trapping potential does not allow for the possibility of trapping 
at the center of the BEC. Taking into account that laboratory BECs possess a finite
lifetime, it is easier to experimentally preserve a soliton at the center of the BEC
where boundary effects are minimal. Therefore we have studied the effect of soliton
trapping for the potential 
\begin{equation}
V(x,t)=ax ^6+f(t-t_o)(\beta x ^4+\gamma x^2)
\label{eq:eq7}
\end{equation}
To obtain experimentally accessible length and time scales for the BEC's size and lifetime
we have used the parameter values $\gamma=0.00398$, $\beta =-1.275 10^{-5}$, 
$\alpha =1.02 10^{-8}$, 
$t_c=100$ and $t_o=950$. After turning on the potential $V_2$ the resulting total potential 
possesses three minima. However the BEC lifetime turns out to be very sensitive with
respect to both the turning on time $t_c$ and the initial properties of the solitonic
excitation. To achieve trapping of the solitonic excitation in the central well of the
total external potential one has to start with a black soliton (i.e. a soliton with zero 
initial velocity). Even small deviations of the initial velocity of the soliton of the order of $10^{-2}$ 
are sufficient to prevent trapping. In Fig.~5(a,b) we show the evolution of a solitonic excitation with initial values 
$(x_o(0),\sin \phi(0))=(-10,0)$ and the external potential given by eq.(\ref{eq:eq7}). 
Fig.~5a shows two snapshots describing the inital state as well as the final state of the composite system. 
We observe the trapping of the initial soliton in the central well of the potential
for times $t > 1000$. BEC radiation effects are 
visible but can be considered as small perturbations for times $t \le 1500$. For 
larger times these perturbations increase substantially finally leading to the destruction of the 
condensate. The corresponding trajectory of the soliton center (solid line) is given in Fig.~5b
together with the results of adiabatic perturbation theory. A very good agreement is also here observed.
This is expected due to the large value of $t_c$ ($t_c=100$). The trapping degree is $m_t \approx 0.56$.

Some additional comments regarding the stability of the BEC are in order. The numerical
simulations reveal that the stability properties of the BEC depend strongly 
on the value for $t_c$ in the switching on function $f(t)$. For small values of $t_c$ the
system becomes very unstable while large values of $t_c$ a significant
stability is achieved. For $t_c \stackrel{>}{\sim} 10$ we encounter a practically stable BEC+soliton system.

\section{Conclusions}

In summary, we have demonstrated that a switching on 
process of a suitably chosen external potential in a quasi one-dimensional BEC leads to trapping
of solitonic excitations in a prescribed subarea of the condensate. 
Depending on the initial soliton's position and velocity
as well as the shape of the external potential the trapping i.e. the confinement of the motion of
the soliton, can be very strong (degree of 
trapping  $m_t \approx 1.0$). For an external potential with two minima there is a finite region
of initial conditions in the phase space of the soliton center for which trapping occurs. 
If the switching on process is slow enough (i.e. of the same order as the 
period of the soliton oscillations) then the corresponding composite system 
(BEC+excitations) is stable for long time intervals and time-independent perturbation thoery predicts
satisfactory the soliton dynamics. The degree of trapping shows a spiral pattern in the phase space of the soliton
which can be understood in terms of the energy transfer during the switching on process.
Using an external potential
with three minima the trapping of the soliton becomes a much more difficult task. Only 
solitons which are initially close to black ones can be trapped in this case;
this can be understood by the fact that shallower solitons have enough initial
kinetic energy to jump the barrier induced by $V_{2}$.
The above results suggest that it is possible to monitor and control solitonic excitations
in BECs by means of proper matter wave devices.

F.K.D. and P.S. acknowledge illuminating discussions with J. Brand.
The work of H.E.N. and D.J.F. has been partially supported by the Special Research Account
of the University of Athens. D.J.F. appreciates the hospitality of the Department of Theoretical 
Chemistry at the University of Heidelberg. We thank also 
G. Theocharis and I. Papacharalampous for their help with respect to computational aspects of
this work. J.S. appreciates support by the DFG Schwerpunktsprogramm 'Interactions in ultracold
Atomic and Molecular Gases'.



{}
\vspace*{1.0cm}

\newpage

\begin{center}
{\Large FIGURES}
\end{center}

\begin{figure}[h]
\scalebox{0.6}{
\rotatebox{270}{
}}
\caption{The contour plot of the trapping degree $m_t$ as a function of
the phase space coordinates $(x_o(0),\sin^2 \phi(0))$ of the initial solitonic 
excitation for the external potential (\ref{eq:eq6}) and $t_c=40$. The calculations
are made within adiabatic perturbation theory.}
\end{figure}

\begin{figure}[h]
\scalebox{0.6}{
\rotatebox{270}{
}}
\scalebox{0.5}{
\rotatebox{270}{
}}
\scalebox{0.5}{
\rotatebox{270}{
}}
\scalebox{0.5}{
\rotatebox{270}{
}}
\caption{Illuminating the soliton reflection process: (a) Three time instants before 
($t=982$), at ($t=1041$) and after ($t=1104$) the reflection of the soliton at the barrier
of the external potential [see eq.(\ref{eq:eq6})], depicted at the trajectory of the soliton center.
The parameters used for the numerical integration of the GPE are the same as in Fig.~1,
for a dark soliton with initial position $x_{o}=-5$ and velocity $\sin\phi(0)=\sqrt{0.15}$.
(b) The density $\vert u(x,t) \vert^2$ and the phase function $\mbox{arg(u)}$ for $t=982$
(before the reflection). (c) The same as in (b), but for $t=1041$ (at the reflection).
Here the density of the soliton becomes minimum and the phase jump across the soliton
becomes step-like. (d) The same as in (b), but for $t=1104$ (after the reflection).
The regions of low and high phases have been interchanged.}
\end{figure}

\begin{figure}[h]
\scalebox{0.7}{
\rotatebox{0}{
}}
\scalebox{0.3}{
\rotatebox{270}{
}}
\caption{(a) The density $\vert u(x,t) \vert^2$ for twelve different times for the 
potential in eq.(\ref{eq:eq6}) and $t_c=40$. The initial conditions for the solitonic
excitation are $(x_o(0),sin\phi(0)) = (-5,\sqrt{0.15})$.
(b) The corresponding trajectory of the center of the solitonic excitation. Both
the numerical GPE integration results (solid line) as well as the adiabatic perturbation
theory results (dashed line) are presented.}
\end{figure}

\begin{figure}[h]
\scalebox{0.7}{
\rotatebox{0}{
}}
\scalebox{0.3}{
\rotatebox{270}{
}}
\caption{(a) The trajectory of the center of the solitonic excitation for the external
potential (\ref{eq:eq6}) and $t_c=15$. The solid line corresponds to the numerical
integration of the GPE while the dashed line is obtained using adiabatic perturbation
theory. The initial conditions for the solitonic excitation are 
$(x_o(0),sin \phi(0)) = (-5,\sqrt{0.15})$.
(b) The same as Fig.~4a with $t_c=70$.}
\end{figure}

\begin{figure}[h]
\scalebox{0.7}{
\rotatebox{0}{
}}
\scalebox{0.3}{
\rotatebox{270}{
}}
\caption{(a) The density $\vert u(x,t) \vert^2$ for two different times $t=200$ and $t=1000$ for the potential in eq.(\ref{eq:eq7}). 
The initial conditions for the solitonic excitation are $(x_o(0),sin \phi(0)) = (-10,0)$.
(b) The corresponding trajectory of the center of the solitonic excitation. Here also
the solid line corresponds to the results obtained by numerical integration of the GPE while the dashed line are the adiabatic perturbation theory results.}
\end{figure}

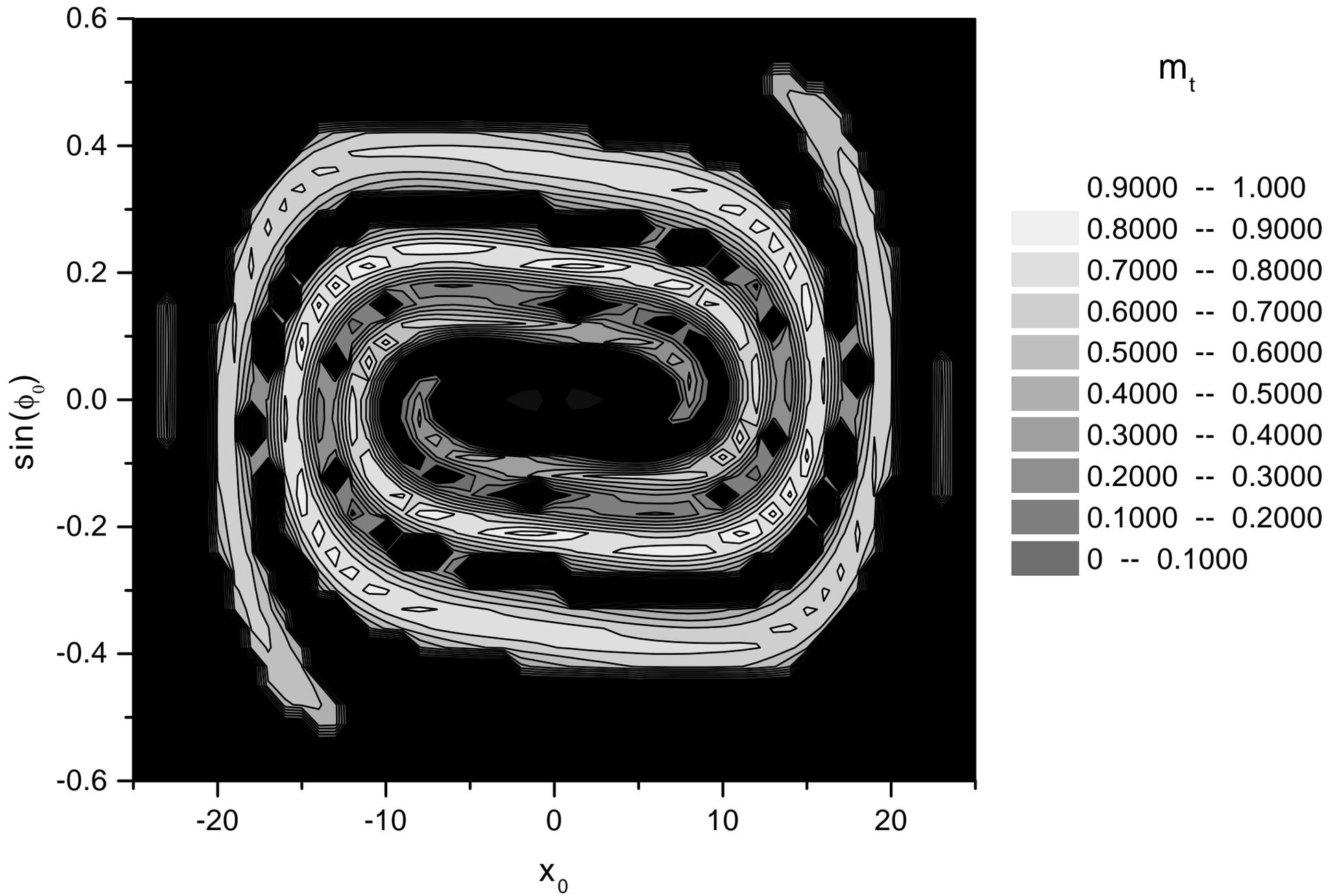
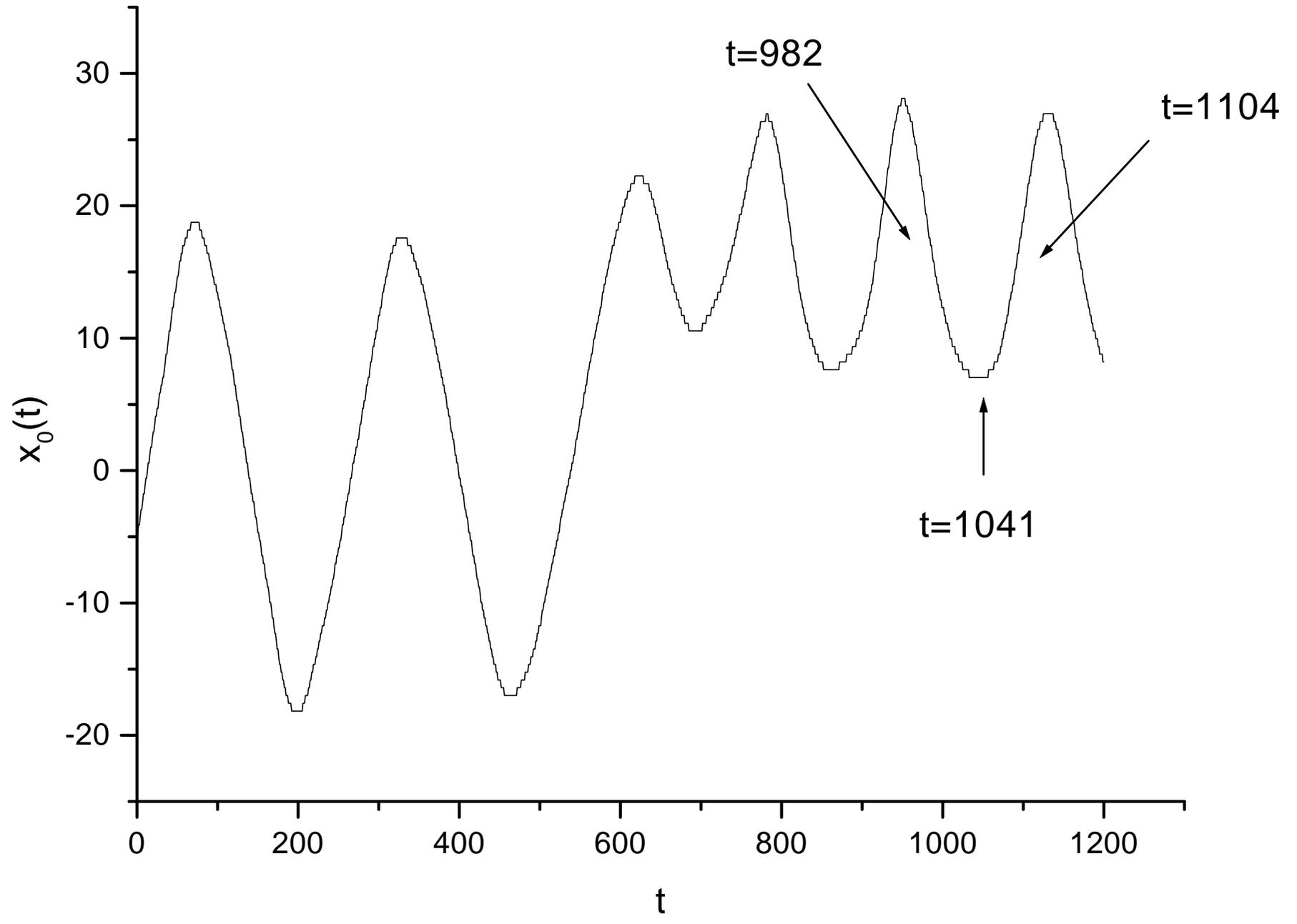
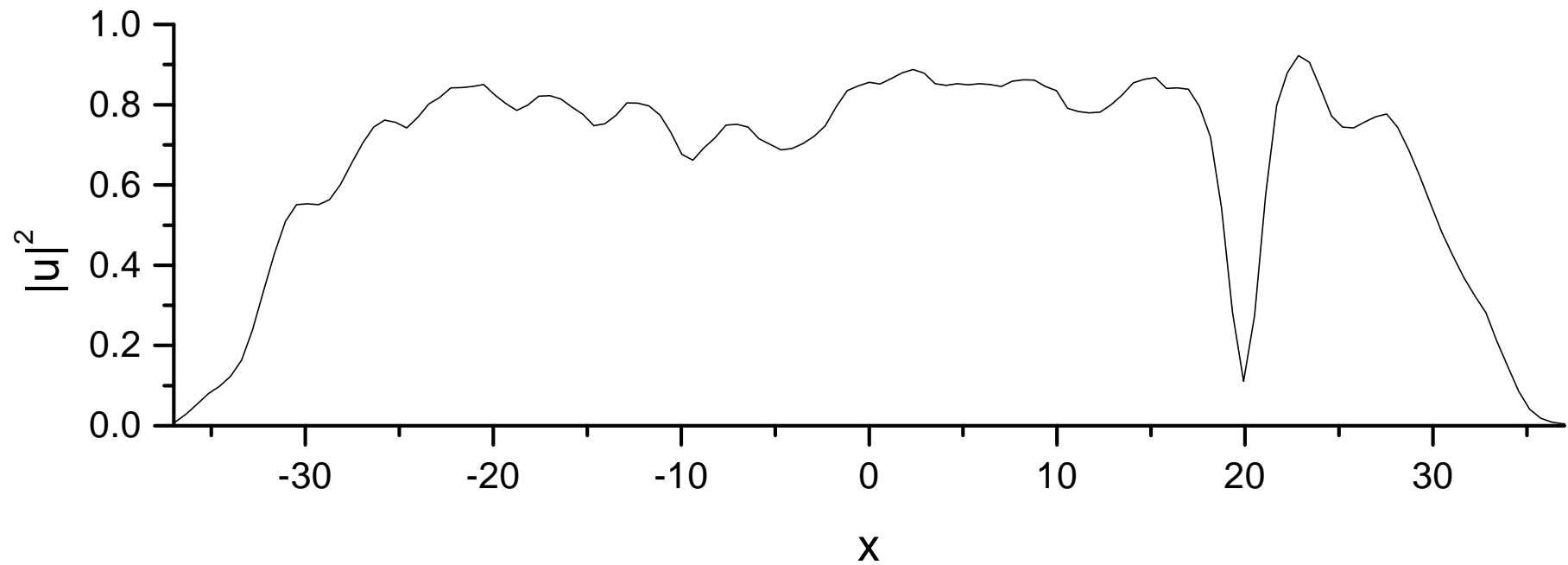
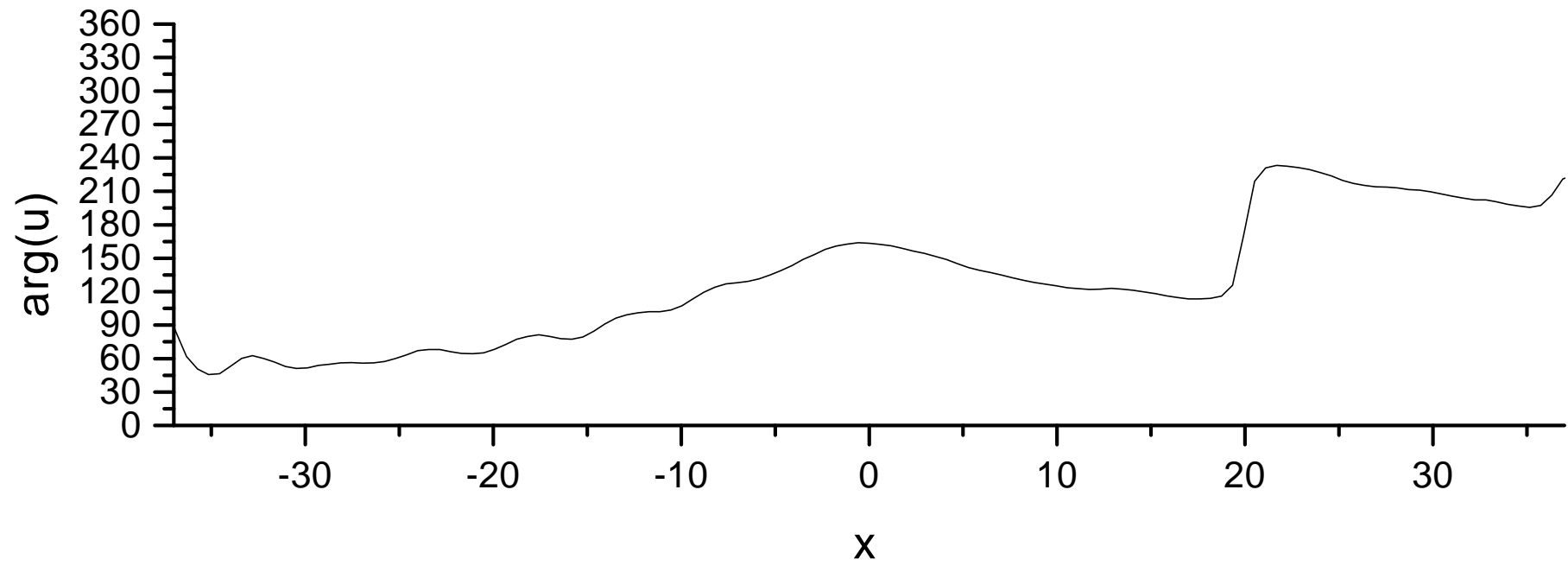
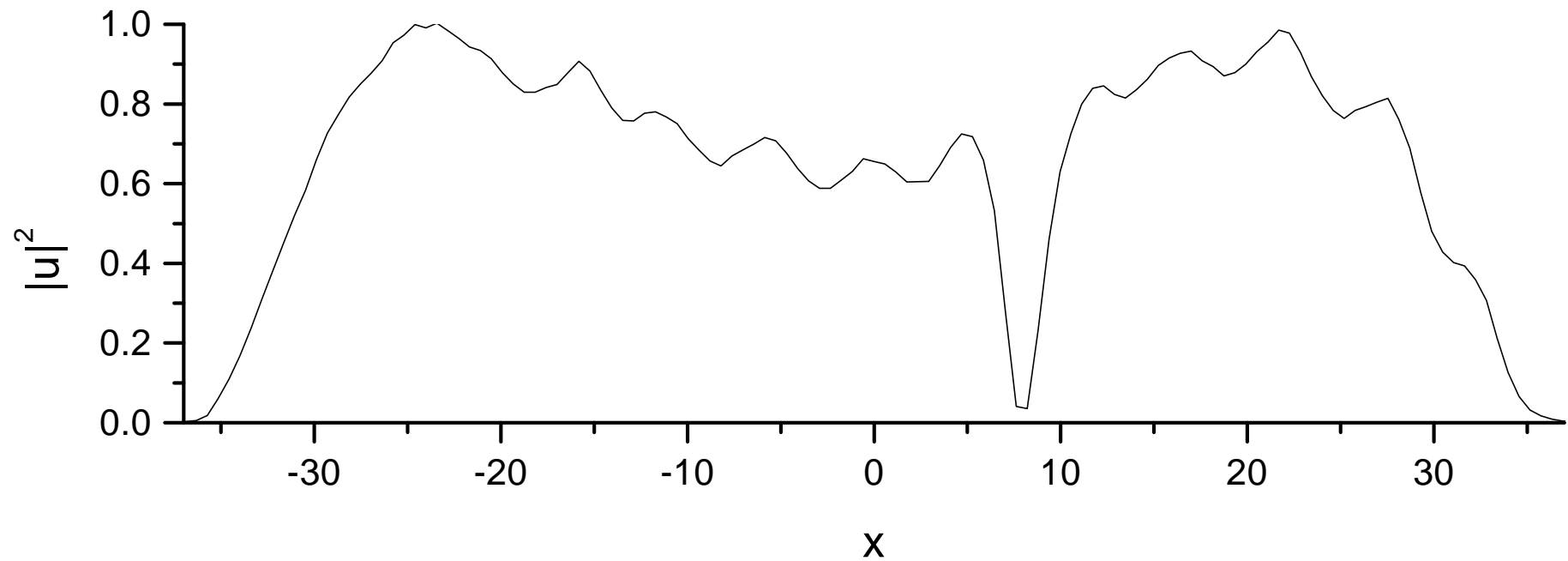
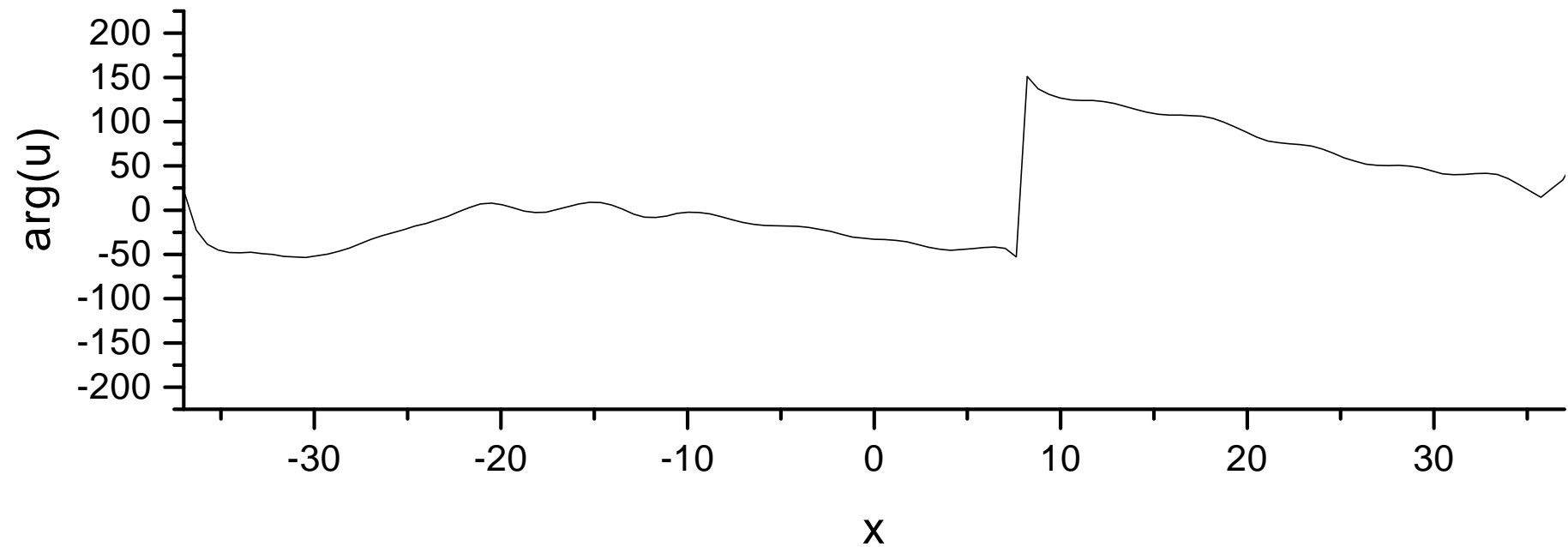
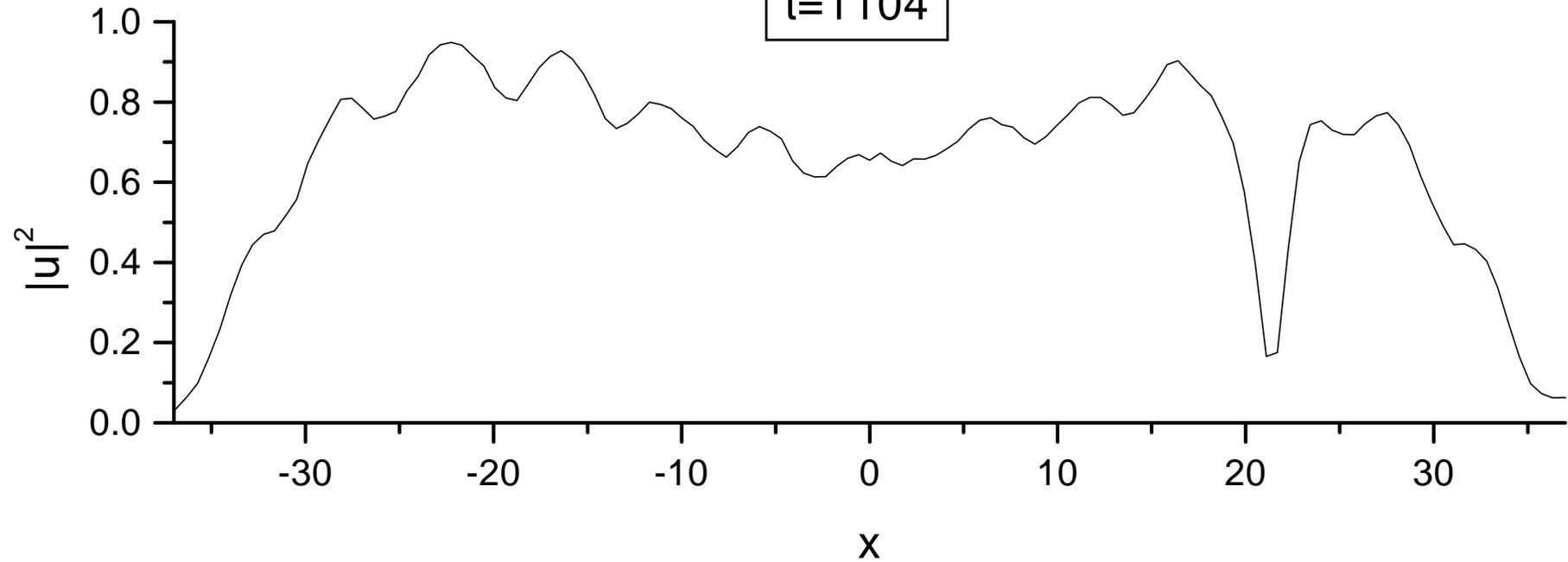
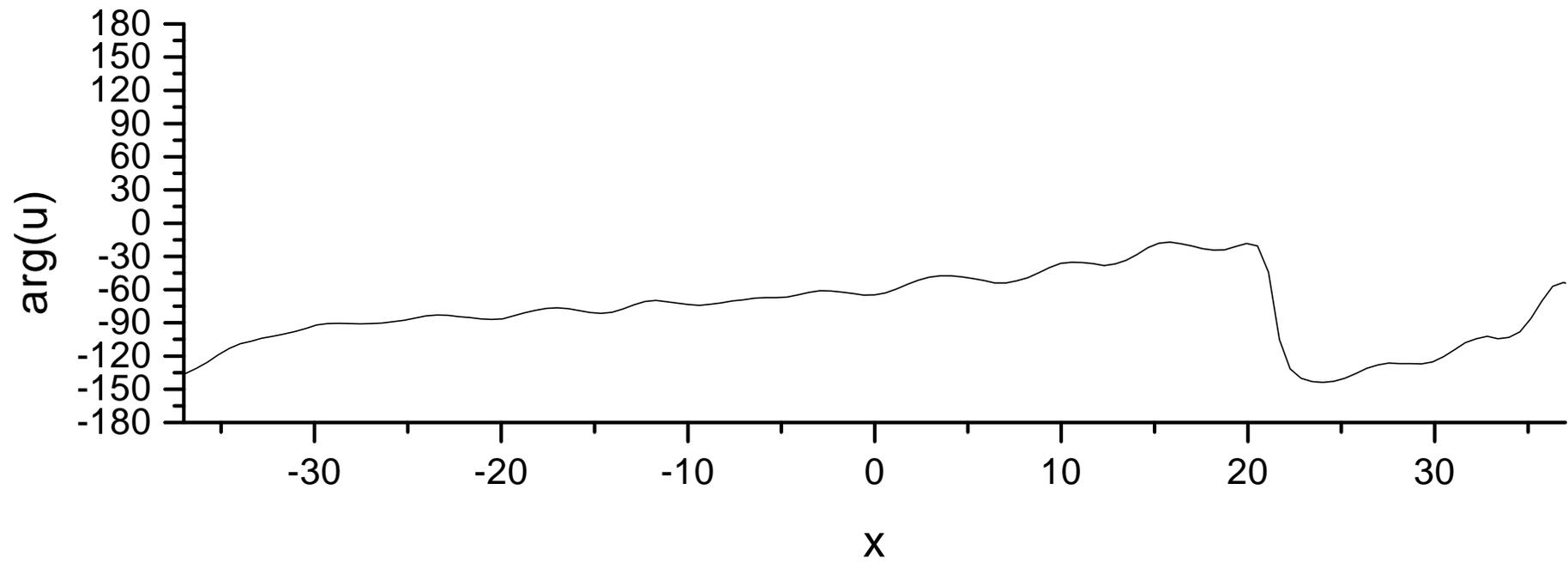
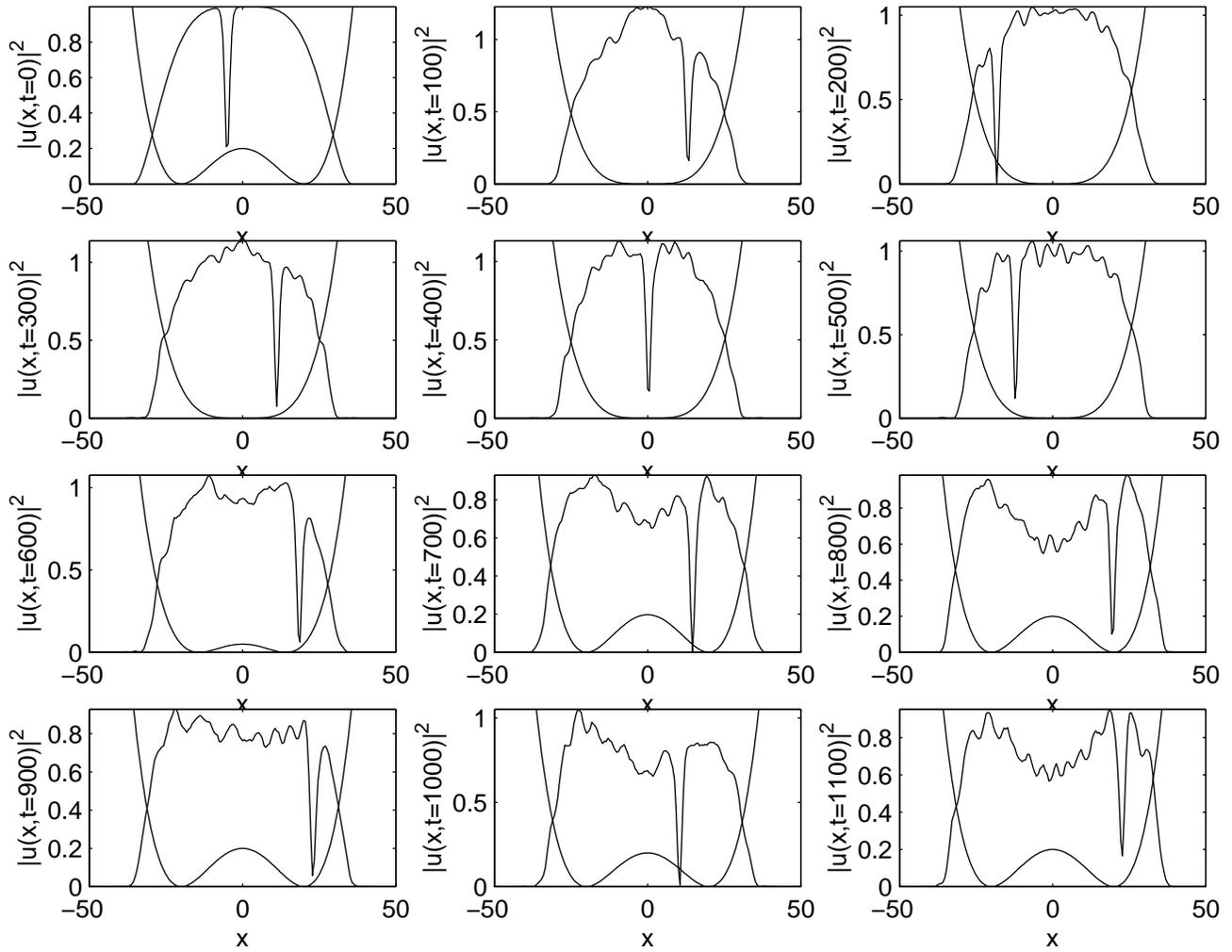
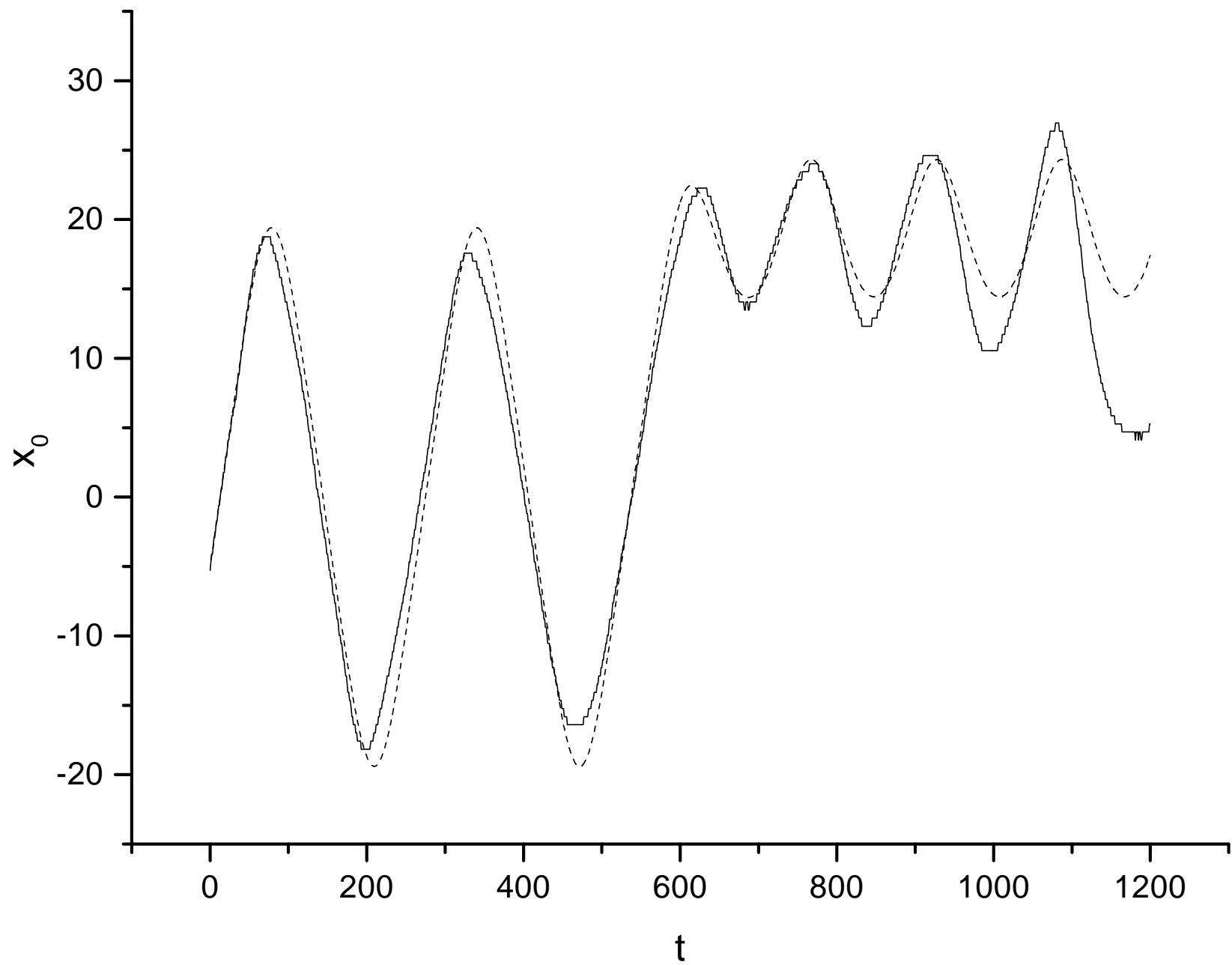
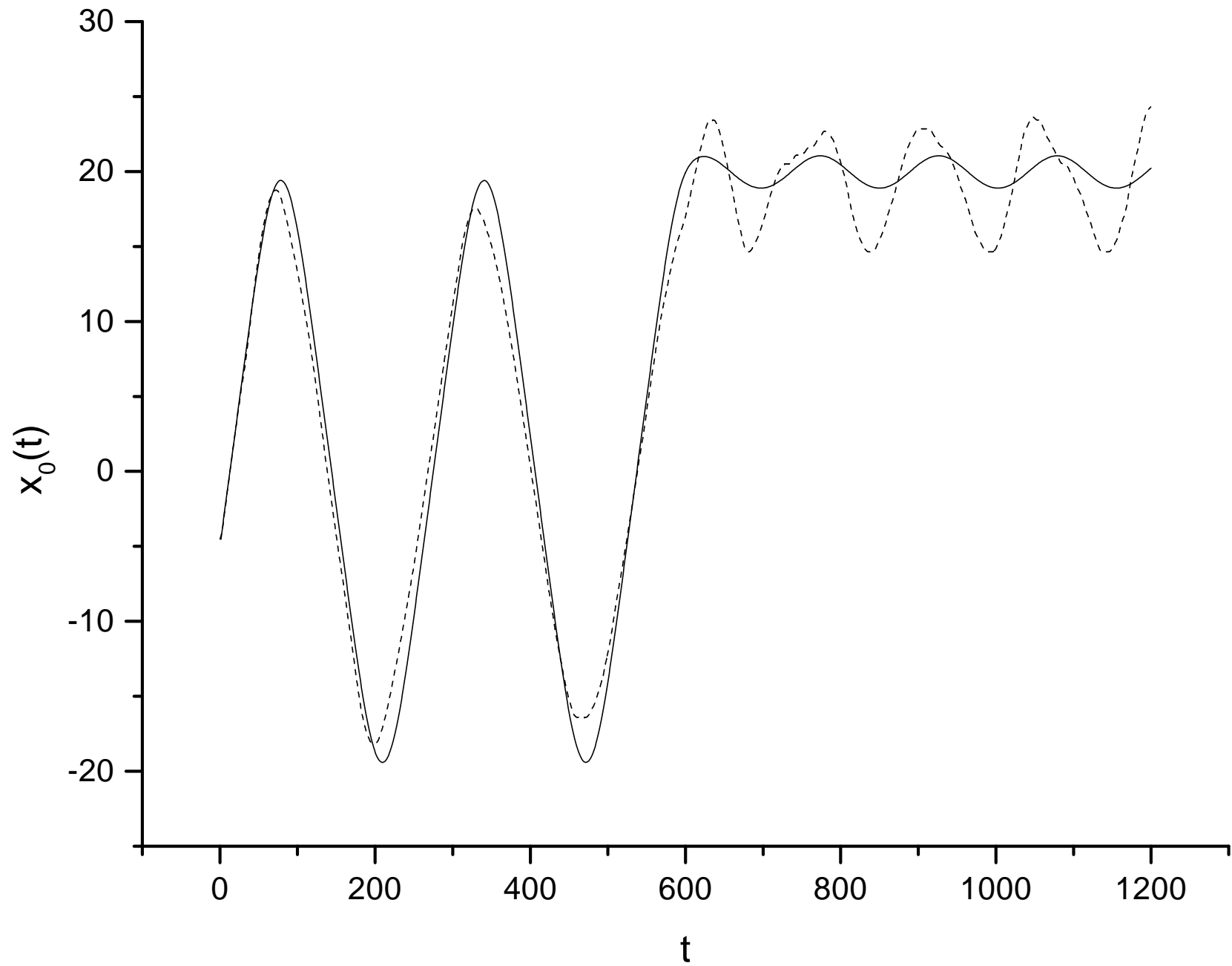
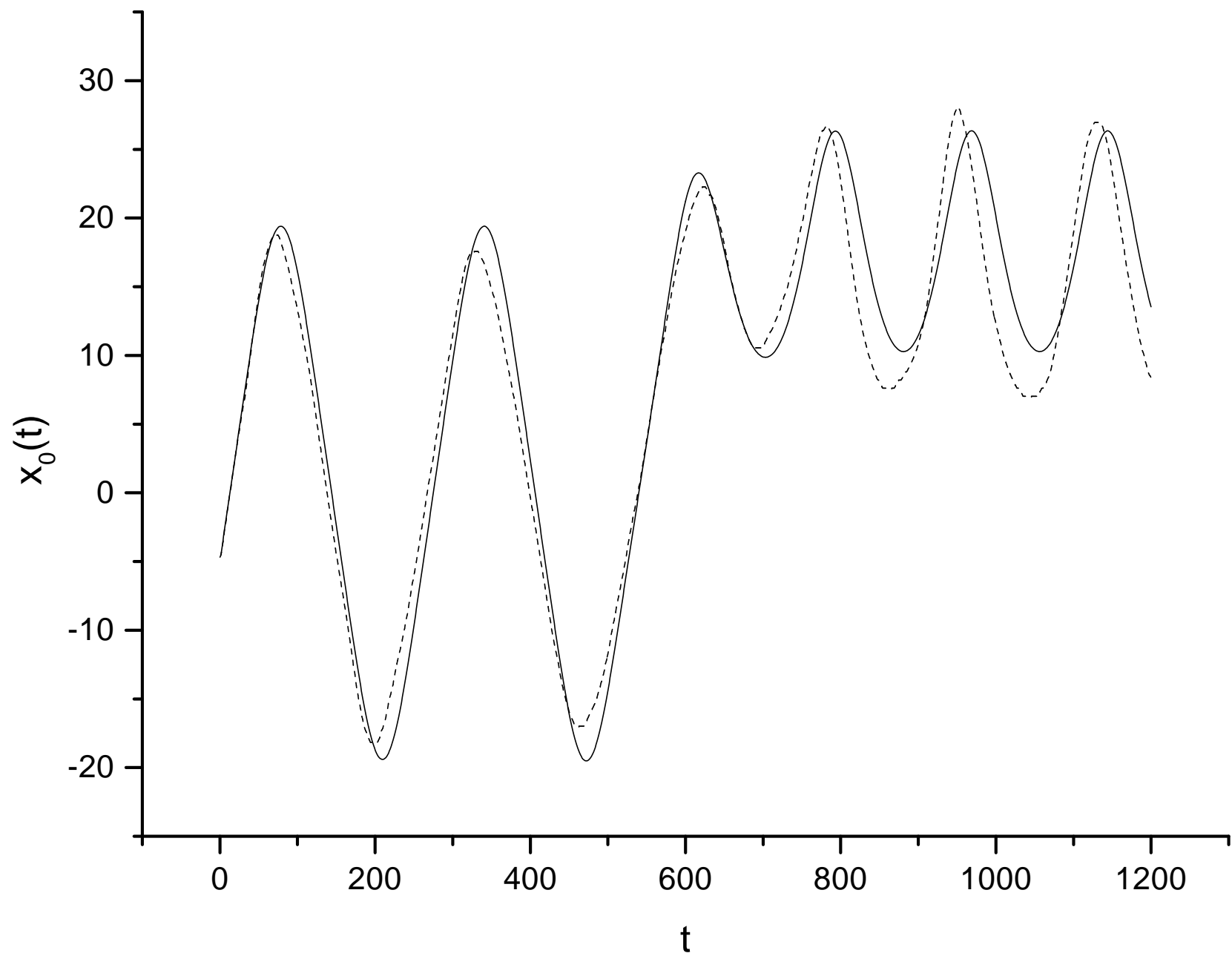
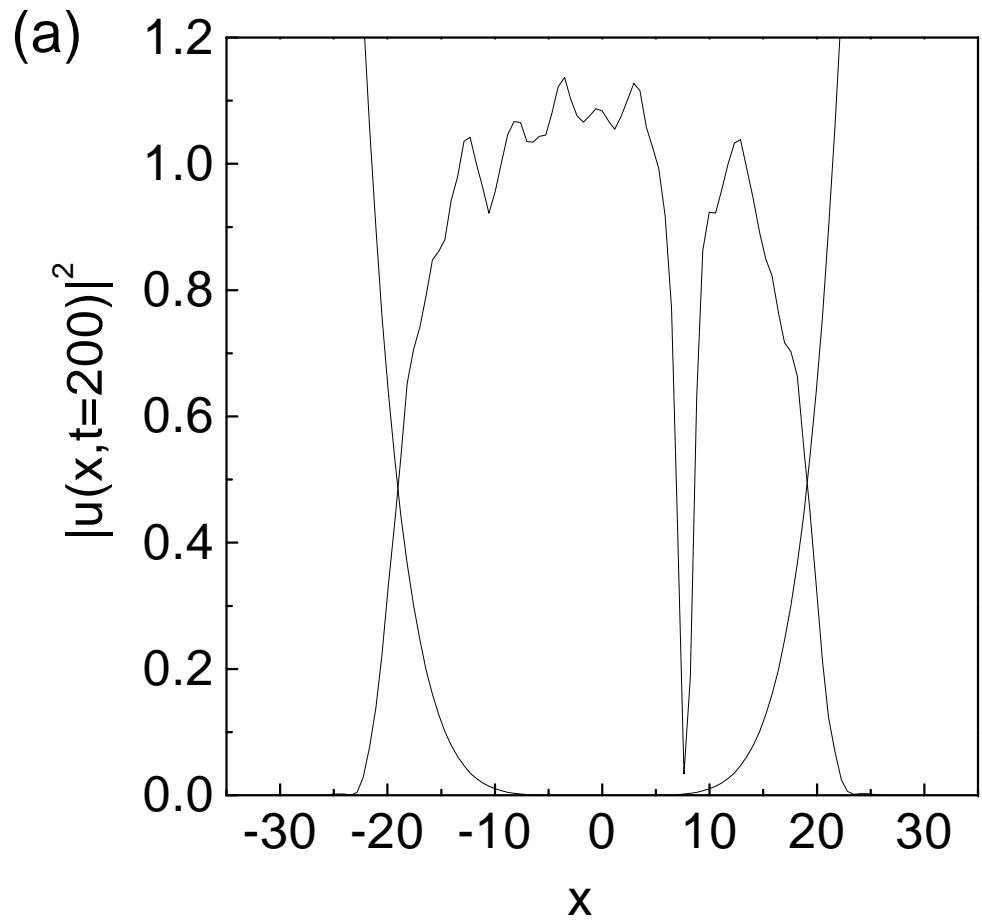
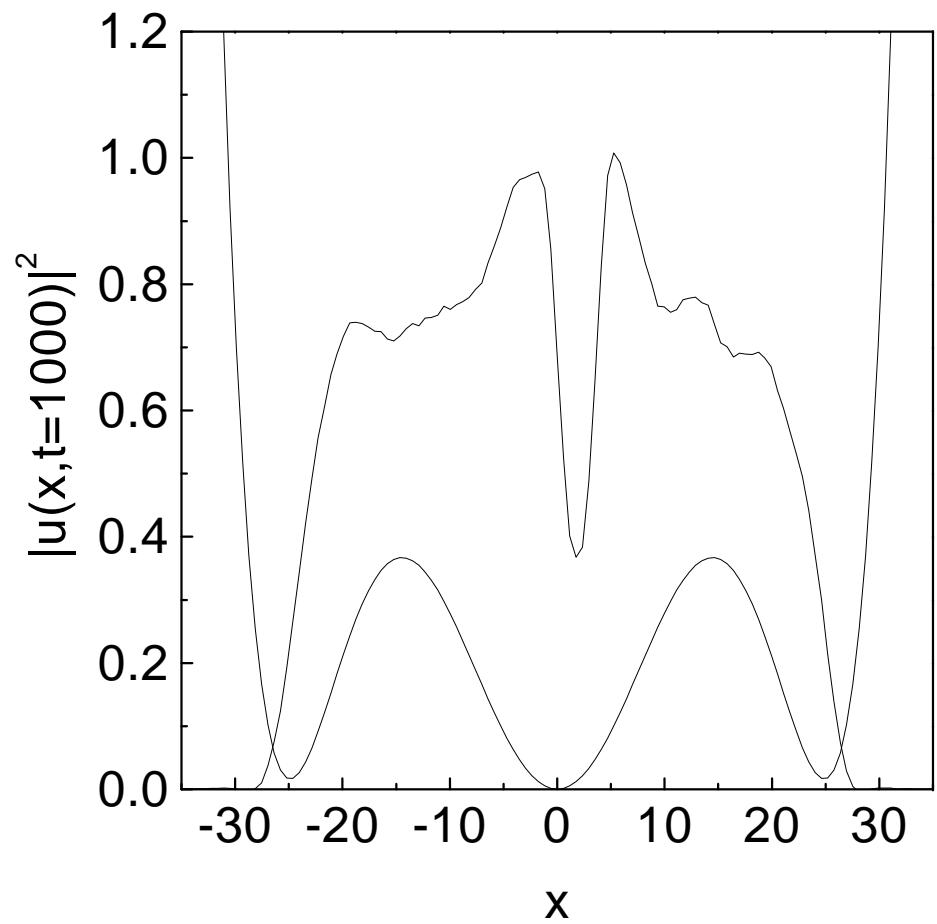
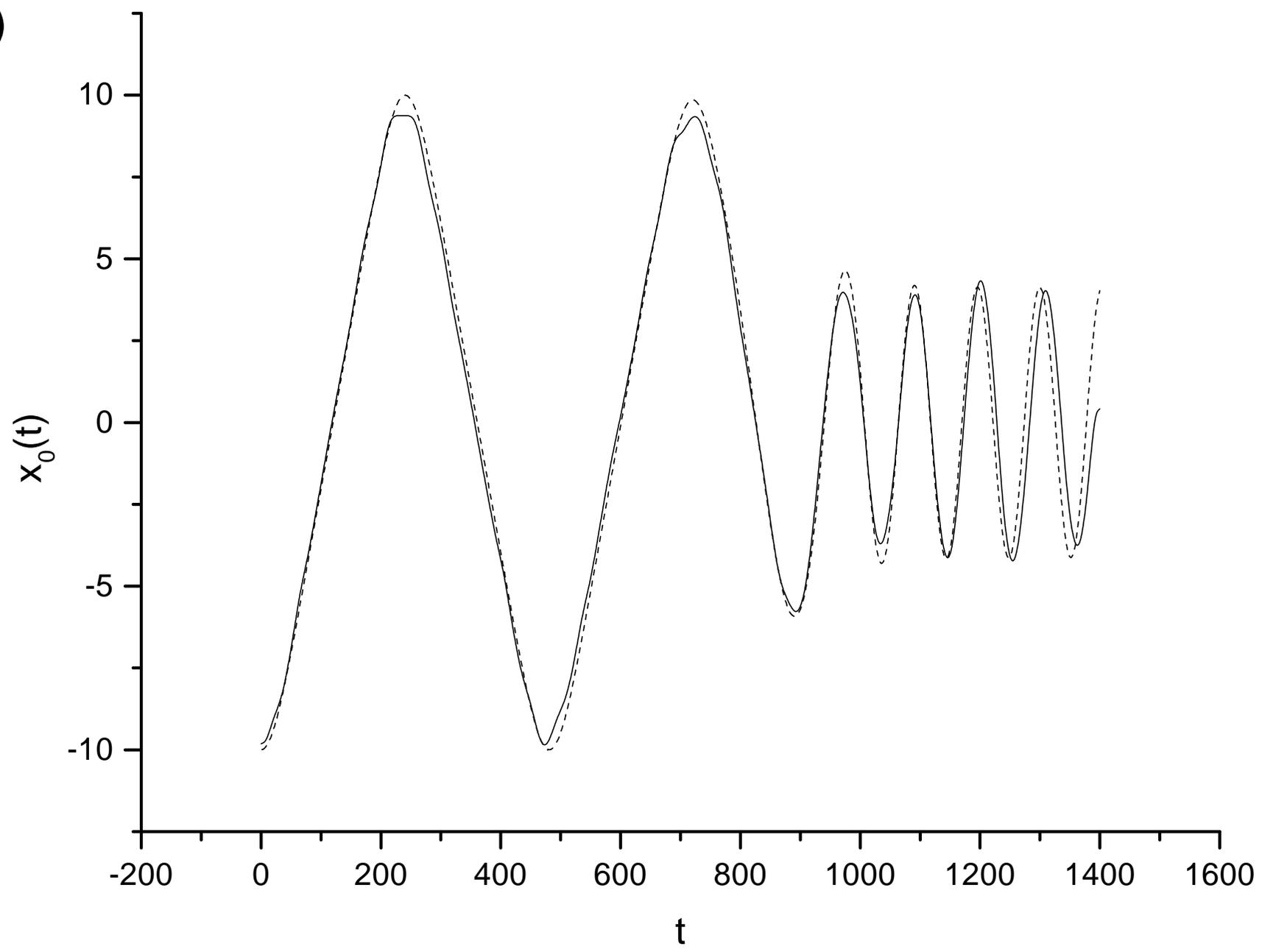
\end{document}